%% file: layerwise_gram_qubo_arxiv.tex
\documentclass[pdflatex,sn-mathphys-num]{sn-jnl}

\usepackage{amsmath,amssymb,amsfonts}
\usepackage{graphicx}
\usepackage{booktabs}
\usepackage{multirow}
\usepackage{algorithm}
\usepackage{algpseudocode}
\usepackage{subcaption}
\usepackage{float}

\usepackage{xcolor}

\makeatletter
\newcommand{\extractorcid}[1]{\@@extractorcid#1\@@nil}
\def\@@extractorcid https://orcid.org/#1\@@nil{#1}
\renewcommand{\orcid}[1]{\unskip\,{\footnotesize\href{#1}{\extractorcid{#1}}}}
\makeatother

\title[Layer-Wise QUBOs]{Layer-wise QUBO-Based Training of CNN Classifiers for Quantum Annealing}

\author*[1,2]{\fnm{Mostafa} \sur{Atallah}\orcid{https://orcid.org/0009-0004-8187-6932}}\email{matalla3@vols.utk.edu}
\author[1]{\fnm{Rebekah} \sur{Herrman}\orcid{https://orcid.org/0000-0001-6944-4206}}

\affil[1]{\orgdiv{Department of Industrial and Systems Engineering},
           \orgname{University of Tennessee - Knoxville},
           \orgaddress{\city{Knoxville}, \postcode{37996}, \state{TN}, \country{USA}}}

\affil[2]{\orgdiv{Department of Physics, Faculty of Science},
           \orgname{Cairo University},
           \orgaddress{\city{Giza}, \postcode{12613}, \country{Egypt}}}

\abstract{Variational quantum circuits for image classification suffer from barren plateaus, while quantum kernel methods scale quadratically with dataset size. We propose an iterative framework based on Quadratic Unconstrained Binary Optimization (QUBO) for training the classifier head of convolutional neural networks (CNNs) via quantum annealing, entirely avoiding gradient-based circuit optimization. Following the Extreme Learning Machine paradigm, convolutional filters are randomly initialized and frozen, and only the fully connected layer is optimized. At each iteration, a convex quadratic surrogate derived from the feature Gram matrix replaces the non-quadratic cross-entropy loss, yielding an iteration-stable curvature proxy. A per-output decomposition splits the $C$-class problem into $C$ independent QUBOs, each with $(d{+}1)K$ binary variables, where $d$ is the feature dimension and $K$ is the bit precision, so that problem size depends on the image resolution and bit precision, not on the number of training samples. We evaluate the method on six image-classification benchmarks (sklearn digits, MNIST, Fashion-MNIST, CIFAR-10, EMNIST, KMNIST). A precision study shows that accuracy improves monotonically with bit resolution, with 10 bits representing a practical minimum for effective optimization; the 15-bit formulation remains within the qubit and coupler limits of current D-Wave Advantage hardware. The 20-bit formulation matches or exceeds classical stochastic gradient descent on MNIST, Fashion-MNIST, and EMNIST, while remaining competitive on CIFAR-10 and KMNIST. All experiments use simulated annealing, establishing a baseline for direct deployment on quantum annealing hardware.
 }


\begin{document}

\renewcommand{\and}{\hskip1em$\cdot$\hskip1em}
\maketitle

\section{Introduction}\label{sec:intro}

Quantum machine learning (QML) seeks to exploit quantum computation for machine learning tasks~\cite{biamonte2017quantum,schuld2015introduction,cerezo2022challenges}. Among the most widely studied applications is image classification, where quantum approaches have been applied through variational quantum classifiers (VQCs)~\cite{schuld2020circuit,grant2018hierarchical}, quantum neural networks~\cite{farhi2018classification,cong2019quantum}, and quantum kernel methods~\cite{singh2022implementation}. VQC-based methods suffer from \emph{barren plateaus}, i.e., exponentially vanishing gradients that make optimization intractable as the number of qubits grows~\cite{mcclean2018barren,cerezo2021cost}. Quantum kernel methods employ parameterized feature-map circuits~\cite{havlicek2019supervised} and face scalability issues since kernel matrix computation requires $O(N^2)$ circuit evaluations for $N$ training samples~\cite{kubler2021inductive}. These limitations motivate alternative paradigms that avoid both gradient-based circuit optimization and quadratic data scaling.

Quadratic unconstrained binary optimization (QUBO) problems map directly to Ising Hamiltonians, the native energy functions of quantum annealers, via a simple linear transformation ($s_i = 2b_i - 1$ converts binary to spin variables)~\cite{lucas2014ising}. Date et al.~\cite{date2021qubo} and Glover et al.~\cite{glover2018tutorial} showed that convex machine learning problems such as linear regression and support vector machines (SVMs) can be cast as QUBOs and solved in a single annealing call. Unlike VQCs, which require estimating gradients of a parameterized quantum circuit, quantum annealing is entirely gradient-free. It encodes the optimization problem in an energy landscape and finds low-energy states through quantum tunneling and thermal fluctuations, so barren plateaus do not arise. Quantum annealing can also escape local minima via tunneling, whereas classical gradient descent may become trapped~\cite{kadowaki1998quantum}.

However, neural network losses are non-convex with multiple local minima~\cite{li2018visualizing}, and the SVM formulation produces an $N\times N$ QUBO matrix that scales with the number of training samples, which is prohibitive for large image-classification datasets. Subi\~{n}as et al.~\cite{subinas2025optimization} proposed an exact QUBO formulation for post-training neural network quantization, encoding rounding decisions as binary variables. Their approach requires pre-trained floating-point weights and one binary variable per weight, making qubit requirements scale with network size.

This work is motivated by the Extreme Learning Machine (ELM) paradigm~\cite{huang2006extreme,park2019convolutional}. Freezing the convolutional filters is essential for our QUBO formulation: because the feature extractor is fixed, the feature matrix $\mathbf{X}$ and the Gram matrix $\mathbf{G} = \frac{1}{N}\mathbf{X}^T\mathbf{X}$ are computed once and remain constant across all iterations, decoupling feature extraction (performed classically) from classifier optimization (performed via QUBO). If the convolutional filters were trainable, every weight update would change $\mathbf{X}$ and invalidate $\mathbf{G}$, requiring the QUBO to be reconstructed at each step and coupling the layers in a non-quadratic optimization. In this work, we freeze randomly initialized convolutional filters and develop a Gram-matrix quadratic surrogate for fully-connected (FC) layer update optimization using softmax residuals and symmetric signed encoding. A layer-wise decomposition splits a $C$-class classifier into $C$ independent per-output QUBOs, each with $(d+1)\times K$ logical binary variables, where $d$ is the feature dimension and $K$ is the bit precision. Notably, this does not depend on the size of the training data, unlike previous approaches. We present an empirical study of bit-precision sensitivity for iterative QUBO training, establishing a minimum precision threshold for effective optimization, and validate the method on six image-classification benchmarks. On the \texttt{sklearn} digits dataset ($8\times 8$ images, $C=10$ classes), our method achieves up to 81.5\% test accuracy with 20-bit precision, compared to 79.8\% for a classical baseline trained under the same frozen-feature setting. Performance improves consistently with precision: 5-bit updates are too coarse (33\% test accuracy), while 10-bit and above produce competitive results. We further validate generalization on five additional benchmarks (MNIST, Fashion-MNIST, CIFAR-10, EMNIST, KMNIST), where QUBO 20-bit matches or exceeds the classical baseline on MNIST (+3\%), Fashion-MNIST (+1.3\%), and EMNIST ($\pm$0\%), while remaining competitive on CIFAR-10 and KMNIST despite the representation bottleneck imposed by $8\times 8$ downsampling.

Our main contributions are:
\begin{enumerate}
\item An iterative Gram-matrix QUBO surrogate that replaces the non-quadratic cross-entropy loss with a convex quadratic amenable to quantum annealing, enabling neural network training from random initialization.
\item A per-output decomposition that reduces QUBO size from $(d{+}1)CK$ to $C$ independent problems of $(d{+}1)K$ variables each, scaling with the model rather than the dataset.
\item An empirical precision-sensitivity study identifying a minimum viable bit precision ($K \geq 10$) for effective QUBO-based training.
\item A multi-dataset benchmark on six image-classification tasks validating the method under a frozen-feature setting.
\end{enumerate}

The remainder of this paper is organized as follows. Section~\ref{sec:related} reviews quantum approaches to image classification. Section~\ref{sec:formulation} develops the QUBO formulation, including the quadratic surrogate, Gram-matrix curvature proxy, and binary encoding. Section~\ref{sec:decomposition} introduces the layer-wise per-output decomposition and presents the training algorithm. Section~\ref{sec:results} reports experimental results on six benchmark datasets, and Section~\ref{sec:discussion} discusses implications and future directions.

\subsection{Related Work}\label{sec:related}
Quantum image classification has been explored through three main paradigms: variational approaches (variational quantum classifiers (VQCs) and quantum convolutional neural networks (QCNNs))~\cite{schuld2020circuit,farhi2018classification,cong2019quantum}, which optimize parameterized quantum circuits via classical gradient loops; non-variational approaches (quantum reservoirs~\cite{kornjaca2024large} and random-circuit methods~\cite{henderson2020quanvolutional}), which use fixed quantum dynamics as feature extractors without gradient optimization; and quantum annealing~\cite{nguyen2019image,schuman2023transfer}, which encodes optimization problems directly in hardware energy landscapes. Each paradigm has distinct strengths and weaknesses. Variational methods offer flexible expressibility but face barren plateaus at scale, non-variational methods avoid gradient issues but sacrifice circuit trainability, and annealing is gradient-free, exploiting quantum tunneling to escape local minima that trap classical optimizers~\cite{kadowaki1998quantum}, but is restricted to problems that can be expressed in QUBO (or equivalently Ising) form, which requires reformulating continuous optimization objectives as quadratic functions of binary variables. We organize the literature accordingly.

Among VQC-based methods for multiclass classification, reported accuracies vary widely depending on the quantum circuit architecture, number of qubits, and degree of classical preprocessing. At the high end of accuracy, hybrid approaches that use classical convolutional neural networks (CNNs) for feature extraction before a small quantum classifier achieve strong results; however, these methods still require gradient-based optimization of the quantum circuit parameters. For example, Senokosov et al.~\cite{senokosov2024quantum} achieve 99.21\% on MNIST ($28\times 28$, 10-class) using parallel 5-qubit VQCs, and Ng et al.~\cite{ng2024hybrid} achieve 94.3\% on MNIST ($10\times 10$, 10-class) with a 6-qubit VQC that encodes images via Hue-Saturation-Value (HSV) color-channel decomposition, which splits each pixel into three perceptually meaningful channels that are independently encoded into qubit rotations. Ensemble approaches such as QUILT~\cite{silver2022quilt} reach $\sim$85\% on MNIST ($28\times 28$, 10-class) using an ensemble of structurally distinct 5-qubit VQCs on IBM hardware. Pure quantum architectures without classical feature extraction achieve lower accuracy: an 8-qubit quantum convolutional neural network (QCNN) with amplitude encoding reaches 57\% on MNIST ($16\times 16$, 10-class)~\cite{mordacci2024multiclass}, outperforming a classical CNN of comparable size but remaining far below hybrid methods. Riaz et al.~\cite{riaz2023nnqe} employed strongly entangled 4-qubit layers, achieving 93.8\% on MNIST ($28\times 28$, 10-class) but only 36\% on CIFAR-10 ($32\times 32$, 10-class). The accuracy gap is not due to the entanglement strategy itself but rather to the increased complexity of CIFAR-10: the classification task remains object recognition in both cases, yet CIFAR-10 images contain three color channels and more complex visual patterns than grayscale handwritten digits, which current few-qubit quantum models struggle to represent. Several VQC-based works report strong results on binary (two-class) tasks, which reduce a $C$-class problem to $\binom{C}{2}$ pairwise one-vs-one classifiers, simplifying each subproblem at the cost of requiring many separate circuits. Liu et al.~\cite{liuao2025lcqhnn} achieved 100\% on MNIST ($28\times 28$, 2-class) and 85.55\% on CIFAR-10 ($32\times 32$, 2-class) with a 4-qubit VQC. Sun et al.~\cite{sun2025sqcnn} reported 99.79\% on MNIST and Fashion-MNIST ($28\times 28$, 2-class) with a scalable QCNN. Hur et al.~\cite{hur2022qcnn} proposed an 8-qubit QCNN achieving up to 99\% on MNIST ($28\times 28$, 2-class). Grant et al.~\cite{grant2018hierarchical} trained hierarchical VQCs on 8 qubits, achieving 99.9\% on MNIST digits 0-vs-1 ($28\times 28$, 2-class). Mari et al.~\cite{mari2020transfer} reached 96.7\% on an ImageNet subset ($224\times 224$, 2-class) via quantum transfer learning. Wilson et al.~\cite{wilson2019quantum} used quantum random feature maps on a Rigetti QPU, achieving 98.6\% on MNIST ($28\times 28$, 2-class, digits 3-vs-5). While these works report strong binary accuracy, they do not address full multiclass classification. A common limitation across all VQC approaches is the reliance on classical gradient-based optimization of circuit parameters, which becomes intractable as qubit counts grow due to barren plateaus~\cite{mcclean2018barren}.

Non-variational quantum approaches sidestep gradient optimization of quantum circuits entirely. Henderson et al.~\cite{henderson2020quanvolutional} introduced quanvolutional neural networks that preprocess MNIST ($28\times 28$, 10-class) images with random 9-qubit quantum circuits as convolutional filters, followed by a classical neural network for classification, reaching $\sim$99\% accuracy. Kornja\v{c}a et al.~\cite{kornjaca2024large} demonstrated quantum reservoir computing on QuEra's Aquila neutral-atom processor, using Rydberg atom dynamics as a fixed nonlinear feature map followed by a classical linear classifier, achieving 93.5\% accuracy on MNIST ($28\times 28$, 2-class) with a 9-qubit chain and scaling to 108 qubits on a 3-class task. In these methods, the quantum circuit is fixed (not trained), so barren plateaus do not arise; however, the quantum component serves only as a feature extractor, with all classification performed by the classical model.

On the quantum annealing side, several works have applied D-Wave hardware to image classification tasks. Nguyen and Kenyon~\cite{nguyen2019image} used D-Wave~2X for sparse coding on MNIST ($12\times 12$, 10-class), achieving 95.68\% accuracy with a linear SVM on the sparse features. Benedetti et al.~\cite{benedetti2017quantum} trained a Boltzmann machine on D-Wave~2X, classifying $7\times 6$ OptDigits images (4-class) at 90\% accuracy. Boyda et al.~\cite{boyda2017deploying} applied QBoost on D-Wave~2X for binary tree-cover detection in aerial imagery, reaching 91.7\%. Koshka and Novotny~\cite{koshka2018comparison} embedded a restricted Boltzmann machine (RBM) on D-Wave~2000Q for binary classification of $8\times 8$ bar patterns, with D-Wave replacing classical sampling for inference. Caldeira et al.~\cite{caldeira2019restricted} trained RBMs on D-Wave~2000Q for binary galaxy morphology classification, finding quantum annealing competitive with classical methods on small datasets. Sleeman et al.~\cite{sleeman2020hybrid} combined a convolutional autoencoder with a quantum RBM on D-Wave~2000Q, achieving 72\% on MNIST ($28\times 28$, 10-class). Asaoka and Kudo~\cite{asaoka2023nbmf} formulated matrix factorization as a QUBO on D-Wave~2000Q for MNIST classification, reaching $\sim$70\% accuracy. These annealing approaches are limited to specific model classes (RBMs, sparse coding, boosting) and do not address direct neural network weight optimization. While some achieve high accuracy on their respective tasks, these specialized formulations do not generalize to training arbitrary neural network architectures, limiting their applicability to more complex classification problems.

A common hybrid strategy across these works is to extract features classically (e.g., using a pretrained CNN such as ResNet) and then train a variational quantum circuit as the classifier head. Mari et al.~\cite{mari2020transfer} formalized this as quantum transfer learning: a classical CNN reduces the input to a low-dimensional feature vector, which is angle-encoded into qubits (one rotation per feature). A parameterized variational circuit processes these qubits, and for a $C$-class problem, the Pauli-$Z$ expectation values of $C$ output qubits are passed through a classical softmax to produce class probabilities. The circuit parameters are optimized via gradient-based methods (e.g., the parameter-shift rule). While this shares our approach's separation of classical feature extraction from quantum classification, these hybrid VQC methods still rely on gradient-based optimization of the quantum circuit parameters and therefore remain susceptible to barren plateaus as the number of qubits grows~\cite{mcclean2018barren}. Multi-class classification further compounds this issue: for $C$ classes, at least $C$ output qubits are needed, and circuit depth must increase to create sufficient entanglement, exacerbating barren plateau severity.

In principle, the classical feature extractor used in such hybrid approaches could also serve as the front end of a QUBO-based classifier: the QUBO formulation is agnostic to the source of the features and depends only on the feature dimension~$d$. However, pretrained networks such as ResNet-18 produce $d=512$ features, yielding per-class QUBOs with $513K$ binary variables, far exceeding the ${\sim}5{,}000$ qubit capacity of current D-Wave Advantage hardware even at moderate bit precision. Our use of a minimal convolutional architecture ($d=18$, giving $19K$ variables per class) is therefore a deliberate design choice that keeps the QUBO within hardware limits, rather than a simplification that could be trivially replaced by a deeper feature extractor. Conversely, if hybrid VQC methods such as by Mari et al. \cite{mari2020transfer} \ adopted frozen random convolutional filters instead of pretrained ResNet features, they would operate on the same low-quality feature space as our method, but would still require gradient-based optimization of the variational circuit parameters (retaining susceptibility to barren plateaus) and would need at least $C$ output qubits with increasing circuit depth for multi-class classification. Under identical frozen-feature conditions, our QUBO-based approach avoids both limitations: optimization is gradient-free, and multi-class scaling requires only $C$ independent fixed-size QUBOs.

Our approach occupies a distinct niche: following the ELM paradigm, convolutional layers are randomly initialized and frozen, and we use QUBO-based iterative optimization to train only the fully connected (FC) classifier layer from random initialization on $8\times 8$ images, achieving 81.3\% on MNIST with 380 logical qubits per QUBO (20-bit precision). Like hybrid VQC methods, features are extracted classically, but instead of training a variational circuit via gradients, our formulation encodes the weight-update optimization as a QUBO solved via annealing, entirely gradient-free and immune to barren plateaus. Multi-class classification scales naturally: each of the $C$ classes yields an independent QUBO of fixed size $(d{+}1)K$, solvable in parallel without increasing circuit depth or qubit count per problem. Images are downsampled to $8\times 8$ to keep the QUBO size within the qubit and coupler limits of current D-Wave hardware. The lower accuracy relative to $28\times 28$ methods reflects this downsampling rather than optimizer limitations, as our QUBO training matches or exceeds a classical baseline under identical frozen-feature conditions.  In contrast to most prior works, which report only accuracy, we also evaluate precision, recall, F1, Cohen's Kappa, and MCC in Section~\ref{sec:benchmark} to provide a more complete assessment.

\section{QUBO Formulation for CNN Training}\label{sec:formulation}

We formulate CNN training for $C$-class classification as a regularized empirical risk minimization problem. Given a training set $\{(\mathbf{x}_n, \mathbf{y}_n)\}_{n=1}^N$ where $\mathbf{x}_n$ is an input image and $\mathbf{y}_n \in \{0,1\}^C$ is a one-hot label vector, we seek network parameters $\boldsymbol{\theta} = \{\mathbf{W}^{(l)}, \mathbf{b}^{(l)}\}_{l=1}^L$ that minimize the regularized cross-entropy loss, $f(\boldsymbol{\theta})$,
\begin{equation}\label{eq:risk}
\min_{\boldsymbol{\theta}} f(\boldsymbol{\theta}) = \frac{\lambda}{2} \|\boldsymbol{\theta}\|_2^2 + \frac{1}{N} \sum_{n=1}^N \ell_n(\boldsymbol{\theta})
\end{equation}
where $\|\boldsymbol{\theta}\|_2^2 = \sum_i \theta_i^2$ is the squared L2 norm over all parameters, $N$ is the number of training samples, the per-sample loss is $\ell_n(\boldsymbol{\theta}) = -\sum_{c=1}^{C} y_{n,c} \log(\hat{y}_{n,c})$ with $\hat{y}_{n,c}$ being the predicted probability for class $c$, and $\lambda > 0$ is the L2 regularization coefficient (a penalty on squared weight magnitudes that discourages overfitting by keeping weights small). We use $\lambda = 0.001$ throughout this work. In our iterative QUBO framework, $\lambda$ controls the magnitude of each weight update: smaller values allow larger steps per iteration, enabling faster convergence, while larger values constrain updates and slow progress. The value $\lambda = 0.001$ is a standard choice in neural network training that balances convergence speed with regularization, and we found it to work well across all benchmarks without further tuning.

This formulation violates both QUBO requirements. First, cross-entropy with softmax normalization is highly nonlinear: the loss involves $\log(\exp(\cdot)/\sum\exp(\cdot))$ compositions that cannot be expressed as a quadratic in the parameters. Second, the loss landscape has multiple local minima, so unlike convex problems where a single QUBO solve finds the global optimum, neural network training requires iterative descent from random initialization. Additionally, the exact Hessian of cross-entropy loss depends on current predictions, changing at every iteration, which would require reformulating the QUBO at each step. We resolve these obstacles by optimizing discrete weight \emph{updates} rather than absolute parameters, replacing the non-quadratic loss with a convex quadratic surrogate, and using iteration to navigate the non-convex landscape.

\subsection{Quadratic Surrogate for Iterative Updates}\label{sec:surrogate}

At each iteration, we seek a weight update $\boldsymbol{\theta}_{t+1} = \boldsymbol{\theta}_t + \boldsymbol{u}$. Rather than optimizing $f(\boldsymbol{\theta}_t + \boldsymbol{u})$ directly, we approximate the loss landscape as a second-order Taylor expansion around the current weights, which is locally quadratic, and construct a quadratic surrogate. This approach is well-established in classical optimization~\cite{lin2020optimization,nocedal2006numerical,bottou2018optimization} but has not been applied in the context of quantum machine learning:
\begin{equation}\label{eq:surrogate}
q(\boldsymbol{u}) = \frac{1}{2} \boldsymbol{u}^T \mathbf{G} \boldsymbol{u} + \mathbf{g}^T \boldsymbol{u}
\end{equation}
where $\mathbf{G} \in \mathbb{R}^{d \times d}$ is a positive semidefinite curvature matrix (an approximation to the Hessian that captures how the loss surface curves near the current weights), $\mathbf{g} \in \mathbb{R}^d$ is the gradient at the current iterate, $d$ is the feature dimension, and $N$ is the number of training samples. Both $\mathbf{G}$ and $\mathbf{g}$ are computed from the training data (see Sections~\ref{sec:gram} and~\ref{sec:gradient}). The matrix $\mathbf{G}$ is positive semidefinite by construction, as shown in Section~\ref{sec:gram}. Newton's method uses the exact Hessian for $\mathbf{G}$, while quasi-Newton methods (BFGS~\cite{nocedal2006numerical}, L-BFGS~\cite{liu1989limited}) use approximations. The key insight is that this surrogate is quadratic in $\boldsymbol{u}$, making it directly suitable for QUBO formulation after binary encoding. Computing $\mathbf{G}$ and $\mathbf{g}$ requires $O(Nd^2)$ and $O(Nd)$ operations respectively (one pass over the training data), comparable to computing gradients for stochastic gradient descent (SGD)~\cite{robbins1951stochastic}. As shown later, our approach will be able to avoid scaling dependence on $N$.

\subsection{The Gram Matrix as Curvature Proxy}\label{sec:gram}

For softmax cross-entropy, the Gauss--Newton matrix provides a curvature approximation that is guaranteed positive semidefinite, unlike the full Hessian which may have negative eigenvalues in non-convex regions~\cite{nocedal2006numerical}. This matrix is $\mathbf{H} = \frac{1}{N}\mathbf{X}^T \mathbf{R} \mathbf{X}$, where $\mathbf{X} \in \mathbb{R}^{N \times d}$ is the feature matrix and $\mathbf{R}$ is a probability-dependent weighting matrix that changes at every iteration. Since $\mathbf{R}$ depends on the current predictions, using $\mathbf{H}$ directly would require recomputing the curvature matrix at every iteration. Inspired by the classical Gram matrix approach used in kernel methods and least-squares optimization~\cite{bishop2006pattern}, we replace $\mathbf{H}$ with the unweighted Gram matrix
\begin{equation}\label{eq:gram}
\mathbf{G} = \frac{1}{N}\mathbf{X}^T\mathbf{X}.
\end{equation}
This matrix is positive semidefinite ($\mathbf{G} \succeq 0$) by construction: for any vector $\mathbf{v}$,
\begin{equation}\label{eq:psd_proof}
\mathbf{v}^T\mathbf{G}\mathbf{v} = \frac{1}{N}\mathbf{v}^T\mathbf{X}^T\mathbf{X}\mathbf{v} = \frac{1}{N}\|\mathbf{X}\mathbf{v}\|_2^2 \geq 0,
\end{equation}
since a squared norm is always non-negative. This is a purely algebraic property of any matrix of the form $\mathbf{A}^T\mathbf{A}$ and does not require any differentiability assumption on the original loss function. The surrogate $q(\boldsymbol{u}) = \frac{1}{2}\boldsymbol{u}^T\mathbf{G}\boldsymbol{u} + \mathbf{g}^T\boldsymbol{u}$ is a quadratic polynomial in $\boldsymbol{u}$, so it is infinitely differentiable with Hessian $\nabla^2 q = \mathbf{G} \succeq 0$. By the second-order convexity condition, a twice-differentiable function is convex if and only if its Hessian is positive semidefinite everywhere~\cite{nocedal2006numerical}; therefore, $q(\boldsymbol{u})$ is convex regardless of the loss landscape. The Gram matrix is also iteration-stable, as it depends only on features $\mathbf{X}$, not on current predictions, and remains constant across iterations when features are frozen. Each entry $G_{ij} = \frac{1}{N}\sum_{n=1}^N x_{ni}\,x_{nj}$ measures the empirical correlation between features $i$ and $j$. Since convolutional features are generally non-orthogonal (neighboring spatial locations and related filters produce correlated activations), most off-diagonal entries $G_{ij}$ are non-zero, making $\mathbf{G}$ a dense matrix. This density propagates to the QUBO matrix $\mathbf{Q}$ (Section~\ref{sec:encoding}), yielding a fully connected optimization problem. Quantum annealers such as D-Wave Advantage have sparse qubit connectivity (the Pegasus graph), so embedding a fully connected QUBO requires chains of physical qubits to represent each logical variable, consuming additional qubits and reducing the effective problem size that the hardware can accommodate (see Section~\ref{sec:discussion}).

Our Gram matrix $\mathbf{G} = \frac{1}{N}\mathbf{X}^T\mathbf{X}$ can be understood as a prediction-independent approximation of the Fisher information matrix~\cite{amari1998natural}, which captures the geometry of the feature space while dropping the output-dependent weighting $\mathbf{R}$. Each QUBO solve thus finds the optimal update with respect to this approximate geometry, making our approach an annealing-based analogue of natural gradient optimization~\cite{ruder2016overview}.

The weighting matrix $\mathbf{R} = \text{diag}(\pi_n(1-\pi_n))$, where $\pi_n \in [0,1]$ is the predicted softmax probability for sample $n$, changes every iteration as predictions evolve. Omitting $\mathbf{R}$ sacrifices curvature accuracy but gains iteration stability: we precompute $\mathbf{G}$ once and reuse it, avoiding $O(Nd^2)$ recomputation per iteration. Since $\pi_n \in [0,1]$, the product $\pi_n(1-\pi_n)$ is maximized at $\pi_n = 0.5$ where it equals $0.25$, so $\mathbf{R}$ acts as a bounded reweighting with entries in $[0, 0.25]$. As predictions approach class certainty ($\pi_n \to 0$ or $\pi_n \to 1$), $\mathbf{R}$ entries shrink toward zero, making the unweighted Gram matrix a reasonable curvature proxy up to constant scaling. Empirically, this approximation works well because the iterative training corrects for curvature inaccuracies over multiple steps. Both $\mathbf{G}$ and $\mathbf{H}$ are dense $(d+1) \times (d+1)$ matrices with the same sparsity pattern, so substituting $\mathbf{G}$ for $\mathbf{H}$ does not increase the QUBO graph density or minor-embedding overhead.

An alternative would be to use a diagonal curvature matrix $\mathbf{G} = \text{diag}(G_{11}, \ldots, G_{(d+1)(d+1)})$, implying a separable QUBO where each parameter can be optimized independently
\begin{equation}\label{eq:diag}
\mathbf{Q} = \mathcal{P}^T \text{diag}(\mathbf{G}) \mathcal{P} = \text{block-diag}\left(G_{11}\mathbf{p}\mathbf{p}^T, \ldots, G_{(d+1)(d+1)}\mathbf{p}\mathbf{p}^T\right).
\end{equation}
While separability simplifies the optimization, it ignores feature correlations that inform how parameters should change together. To capture parameter interactions and expose precision sensitivity, we use the full (dense) Gram matrix.

\subsection{Gradient from Softmax Residuals}\label{sec:gradient}

The gradient $\mathbf{g}$ is the standard calculus derivative of the cross-entropy loss with respect to the weights. For a linear softmax classifier, this derivative has a well-known closed-form expression in terms of the prediction residuals~\cite{bishop2006pattern,goodfellow2016deep}. Let $\mathcal{L} = \frac{1}{N}\sum_n \ell_n$ denote the average cross-entropy loss, which requires one forward pass through the network to compute. Then
\begin{equation}\label{eq:gradient}
\mathbf{g} = \frac{\partial \mathcal{L}}{\partial \boldsymbol{\theta}} = \frac{1}{N}\mathbf{X}^T(\boldsymbol{\pi} - \mathbf{y}) = -\frac{1}{N}\mathbf{X}^T\mathbf{r}
\end{equation}
where $\boldsymbol{\pi} = \text{softmax}(\mathbf{X}\boldsymbol{\theta}) \in [0,1]^N$ is the vector of predicted class probabilities (the $n$-th entry $\pi_n$ is the same per-sample predicted probability used in the weighting matrix $\mathbf{R}$ above), $\mathbf{y}$ are the one-hot targets, and $\mathbf{r} = \mathbf{y} - \boldsymbol{\pi}$ is the softmax residual.

Including L2 regularization on the weights (but not biases), the surrogate becomes
\begin{equation}\label{eq:surrogate_reg}
q_\lambda(\boldsymbol{u}) = \frac{1}{2} \boldsymbol{u}^T \mathbf{G}_\lambda \boldsymbol{u} + \mathbf{g}_\lambda^T \boldsymbol{u}
\end{equation}
where $\mathbf{G}_\lambda = \mathbf{G} + \lambda \cdot \text{diag}([1,\ldots,1,0])$ excludes bias from regularization, and $\mathbf{g}_\lambda = \mathbf{g} + \lambda[\boldsymbol{\theta}_{\text{weights}}; 0]$ adds the regularization gradient (here $\lambda$ is the same regularization coefficient as in Equation~\eqref{eq:risk}). Each individual surrogate is convex in $\boldsymbol{u}$ for fixed $\mathbf{G}_\lambda \succeq 0$; however, the composition of repeated updates under a nonlinear softmax makes the overall training procedure non-convex. The quadratic surrogate approximates the local loss landscape at each step, while iteration handles the global non-convexity: after each update, the gradient $\mathbf{g}$ is recomputed at the new weights, so successive surrogates explore different regions of the loss surface, analogous to how gradient descent navigates non-convex landscapes through repeated local steps.

\subsection{Binary Encoding}\label{sec:encoding}

Quantum annealers optimize unconstrained binary quadratic functions, so we discretize continuous updates using symmetric signed encoding~\cite{date2021qubo}. We define a precision vector $\mathbf{p} = [p_0, \ldots, p_{K-1}]^T \in \mathbb{R}^K$ with powers-of-2 encoding
\begin{equation}\label{eq:precision}
p_k = \frac{\delta_{\max}}{2^K - 1} \cdot 2^k, \quad k = 0, \ldots, K-1
\end{equation}
where $\delta_{\max} > 0$ is the maximum update magnitude, $K$ is the bit precision, and $k$ is a summation index running over the $K$ bit positions. The update vector $\boldsymbol{u}$ from Section~\ref{sec:surrogate} is encoded component-wise: for each parameter index $j \in \{1, \ldots, d{+}1\}$, the $j$-th component $u_j$ is represented by a binary sub-vector $\mathbf{b}_j \in \{0,1\}^K$, where $b_{j,k}$ denotes the $k$-th bit of the $j$-th parameter's encoding,
\begin{equation}\label{eq:encoding}
u_j = \sum_{k=0}^{K-1} p_k (2b_{j,k} - 1) = \mathbf{p}^T(2\mathbf{b}_j - \mathbf{1}).
\end{equation}
The $(2b-1)$ mapping centers updates around zero, with range $[-\delta_{\max}, +\delta_{\max}]$. The resolution is $p_0 = \delta_{\max}/(2^K-1)$, and all parameters share the same precision vector. We set $\delta_{\max} = 0.5$ throughout this work. This value was chosen empirically to allow meaningful weight changes per iteration while preventing instability from oversized updates. We conducted preliminary experiments with $\delta_{\max} \in \{0.1, 0.25, 0.5, 1.0\}$ and found $\delta_{\max} = 0.5$ to balance convergence speed with stability across datasets.

For example, when $\delta_{\max}=0.5$, $K=5$ gives resolution $\approx 0.016$, while $K=20$ yields resolution $\approx 4.8 \times 10^{-7}$. This encoding does not guarantee that $u_j = 0$ is exactly representable for all $K$; the closest value to zero has magnitude $\epsilon(K) = p_0 = \delta_{\max}/(2^K-1)$. For $K=10$ and $\delta_{\max}=0.5$, $\epsilon \approx 4.9 \times 10^{-4}$; for $K=20$, $\epsilon \approx 4.8 \times 10^{-7}$. These errors are negligible compared to typical weight magnitudes ($\sim 0.1$--$1.0$).

\subsection{QUBO Matrix Construction}\label{sec:qubo_matrix}

Substituting the binary encoding into the quadratic surrogate $q_\lambda(\boldsymbol{u})$ from Equation~\eqref{eq:surrogate_reg} yields the QUBO. With $\mathbf{u} = 2\mathcal{P}\mathbf{b} - \delta_{\max}\mathbf{1}_{d+1}$ where $\mathbf{1}_{d+1} \in \mathbb{R}^{d+1}$ is the all-ones vector and $\mathcal{P} = \mathbf{I}_{d+1} \otimes \mathbf{p}^T \in \mathbb{R}^{(d+1) \times (d+1)K}$ is a block-diagonal matrix formed by the Kronecker product of the $(d{+}1)$-dimensional identity matrix with the precision vector, the QUBO energy (up to constant terms that do not affect the minimizer) is
\begin{equation}\label{eq:qubo_energy}
E(\mathbf{b}) = \mathbf{b}^T \mathbf{Q} \mathbf{b} + \mathbf{q}^T \mathbf{b}
\end{equation}
where the quadratic and linear coefficients are
\begin{align}\label{eq:qubo_coeffs}
\mathbf{Q} &= 4\mathcal{P}^T \mathbf{G}_\lambda \mathcal{P} \\
\mathbf{q} &= 4\mathcal{P}^T (\mathbf{g} - \mathbf{G}_\lambda \delta_{\max} \mathbf{1}_{d+1}).
\end{align}
The factor of 4 arises from the $2\mathcal{P}\mathbf{b}$ term in the encoding. This constant scaling does not change which binary vector $\mathbf{b}^*$ minimizes the QUBO; only the energy value at that minimum is scaled.

The quadratic terms, $\mathbf{Q}$, capture feature correlations from the Gram matrix, determining how parameters should change together. The linear terms, $\mathbf{q}$, encode the gradient direction (the direction of steepest descent in the loss landscape), indicating which binary configurations reduce the loss. Before solving, we normalize QUBO coefficients by dividing by the maximum absolute coefficient. This does not change the theoretical minimizer since QUBO minimization depends only on relative magnitudes. Empirically, normalization improves solver stability by ensuring consistent annealing temperature schedules across iterations~\cite{kirkpatrick1983optimization}. Without normalization, coefficient magnitudes vary widely across iterations, requiring per-iteration temperature tuning.

\section{Layer-Wise and Per-Class Decomposition}\label{sec:decomposition}

A multi-class classifier does not require a single monolithic QUBO. Let $d$ denote the feature dimension, $C$ the number of classes, and $K$ the bit precision. A full fully connected (FC) layer QUBO would have $(d+1) \times C \times K$ binary variables, which is potentially too large for current quantum hardware. By decomposing the classifier head across output neurons, we obtain $C$ smaller dense QUBOs of size $(d+1) \times K$ each. This decomposition sacrifices cross-class curvature information, since the softmax couples all outputs. This is potentially relevant for datasets with correlated classes such as Fashion-MNIST, where visually similar categories (e.g., Shirt, T-shirt, Pullover, Coat) share feature representations, or CIFAR-10, where automobile and truck share similar visual features. However, the decomposition preserves within-class feature correlations via the shared Gram matrix. The total parameter count remains unchanged; only the optimization is decomposed. Empirically, accuracy is not significantly affected because iterative training corrects for approximation errors over multiple steps.

\subsection{Problem Setup}\label{sec:setup}

Consider a CNN with a frozen feature extractor (i.e., convolutional layers whose weights are randomly initialized and held fixed, as in the ELM paradigm) followed by a fully connected classification layer. The FC layer weights are initialized randomly and serve as the initial state for optimization. No special data encoding is required as the QUBO is constructed from classical features and for quantum annealing deployment, each binary variable maps to a spin variable via $s_i = 2b_i - 1$, with the annealer preparing the corresponding Ising Hamiltonian. Let $N$ denote the number of training samples, $d$ the dimension of flattened CNN features (after pooling), $C$ the number of output classes, and $K$ the bit precision (binary variables per continuous parameter). The FC layer has weight matrix $\mathbf{W} \in \mathbb{R}^{d \times C}$ and bias vector $\mathbf{b} \in \mathbb{R}^C$, totaling $P = (d+1) \times C$ parameters when including biases.

We augment the feature matrix to jointly optimize weights and biases within a single QUBO formulation. This standard technique~\cite{bishop2006pattern} absorbs the bias into the weight vector, eliminating the need for separate bias handling. Let $\mathbf{X} \in \mathbb{R}^{N \times d}$ be the flattened CNN features. The augmented feature matrix appends a column of ones (so that multiplying by the augmented weight vector $[\mathbf{w}; b]$ produces $\mathbf{X}\mathbf{w} + b\mathbf{1}_N$, effectively incorporating the bias). The subscript $N$ on $\mathbf{1}_N$ indicates an $N$-dimensional vector of ones (one entry per training sample)
\begin{equation}\label{eq:augment}
\mathbf{X}_{\text{aug}} = [\mathbf{X}, \mathbf{1}_N] \in \mathbb{R}^{N \times (d+1)}.
\end{equation}
The augmented weight vector for class $c$ combines the $c$-th column of $\mathbf{W}$ and the $c$-th entry of $\mathbf{b}$: $\mathbf{w}_c^{\text{aug}} = [\mathbf{w}_c; b_c] \in \mathbb{R}^{d+1}$, and the full augmented weight matrix is $\mathbf{W}_{\text{aug}} = [\mathbf{w}_1^{\text{aug}}, \ldots, \mathbf{w}_C^{\text{aug}}] \in \mathbb{R}^{(d+1) \times C}$.

\subsection{Per-Output Decomposition}\label{sec:per_output}

We decompose the global optimization into $C$ independent per-output subproblems, one QUBO for each output neuron's weights and bias. The Gram matrix $\mathbf{G} = \frac{1}{N}\mathbf{X}_{\text{aug}}^T\mathbf{X}_{\text{aug}} \in \mathbb{R}^{(d+1) \times (d+1)}$ is shared across all $C$ problems. Although softmax couples outputs across classes (changing one class's weights affects all predicted probabilities), we treat per-class updates independently within each iteration. This corresponds to a block-coordinate update over the columns of $\mathbf{W}_{\text{aug}}$ where each block (column $c$) is optimized using the shared curvature proxy $\mathbf{G}_\lambda$ while holding the remaining blocks fixed, yielding a block-diagonal approximation in class space. This approximation works well empirically because subsequent iterations correct for any suboptimality, and the shared Gram matrix preserves within-class feature geometry.

Section~\ref{sec:formulation} derived the surrogate for a single output; here we instantiate it for each class $c$ independently. For output class $c \in \{1, \ldots, C\}$, the per-class softmax residual specializes Equation~\eqref{eq:gradient} to a single output column
\begin{equation}\label{eq:residual}
\mathbf{r}_c = \mathbf{y}_c - \boldsymbol{\pi}_c \in \mathbb{R}^N \quad \text{(target minus prediction)}
\end{equation}
where $\mathbf{y}_c \in \{0,1\}^N$ is the one-hot target vector and $\boldsymbol{\pi}_c$ is the predicted probability for class $c$. Applying the quadratic surrogate from Section~\ref{sec:surrogate} to this per-class residual yields the per-output optimization problem
\begin{equation}\label{eq:per_output_surrogate}
\min_{\mathbf{u}_c^{\text{aug}} \in \mathbb{R}^{d+1}} \frac{1}{2}(\mathbf{u}_c^{\text{aug}})^T \mathbf{G}_\lambda \mathbf{u}_c^{\text{aug}} + \mathbf{g}_c^T \mathbf{u}_c^{\text{aug}}
\end{equation}
where $\mathbf{G}_\lambda = \mathbf{G} + \lambda \cdot \text{diag}(\underbrace{1,\ldots,1}_{d},0)$ applies L2 regularization to weights but not bias, and the gradient is
\begin{equation}\label{eq:per_output_gradient}
\mathbf{g}_c = -\frac{1}{N}\mathbf{X}_{\text{aug}}^T \mathbf{r}_c + \lambda [\mathbf{w}_{c}; 0] \in \mathbb{R}^{d+1}.
\end{equation}

For symmetric signed encoding $\mathbf{u}_c^{\text{aug}} = 2\mathcal{P}\mathbf{b}_c - \delta_{\max}\mathbf{1}_{d+1}$ where $\mathcal{P} = \mathbf{I}_{d+1} \otimes \mathbf{p}^T \in \mathbb{R}^{(d+1) \times (d+1)K}$, the QUBO energy is
\begin{equation}\label{eq:per_output_qubo}
E(\mathbf{b}_c) = \mathbf{b}_c^T \mathbf{Q}_c \mathbf{b}_c + \mathbf{q}_c^T \mathbf{b}_c + \frac{1}{2}\delta_{\max}^2 \mathbf{1}_{d+1}^T \mathbf{G}_\lambda \mathbf{1}_{d+1} - \delta_{\max} \mathbf{g}_c^T \mathbf{1}_{d+1}
\end{equation}
where
\begin{align}\label{eq:per_output_coeffs}
\mathbf{Q}_c &= 4\mathcal{P}^T \mathbf{G}_\lambda \mathcal{P} \in \mathbb{R}^{(d+1)K \times (d+1)K} \\
\mathbf{q}_c &= 4\mathcal{P}^T (\mathbf{g}_c - \mathbf{G}_\lambda \delta_{\max} \mathbf{1}_{d+1}) \in \mathbb{R}^{(d+1)K}.
\end{align}
Note that $\mathbf{Q}_c$ is identical across all classes (since it depends only on the shared Gram matrix), while $\mathbf{q}_c$ differs per class through the class-specific gradient $\mathbf{g}_c$. The constant terms do not affect the binary minimizer and are dropped during optimization.
This yields $C$ independent dense QUBOs, each with $(d+1) \times K$ binary variables. Each per-output QUBO has $\frac{(d+1)K \cdot ((d+1)K - 1)}{2}$ quadratic terms. The Gram matrix $\mathbf{G} = \frac{1}{N}\mathbf{X}_{\text{aug}}^T\mathbf{X}_{\text{aug}}$ is dense, creating $O((d+1)^2 K^2)$ non-zero entries. Since the convolutional features are frozen, $\mathbf{G}$ is computed once and reused across all iterations, shared among all $C$ per-output QUBOs. After solving each QUBO, the proposed update is applied with step size $\alpha = 1$ (full update). All QUBO solutions are applied unconditionally without acceptance tests, relying on the quality of the surrogate to produce beneficial updates on average.

The feature dimension $d$ is fully determined by the convolutional architecture. For an input image of size $H \times W$ processed by a single convolutional layer with kernel size $k \times k$ (stride~1, no padding) followed by $s \times s$ max pooling, the pooled spatial dimensions are $h_{\text{pool}} = \lfloor (H - k + 1)/s \rfloor$ and $w_{\text{pool}} = \lfloor (W - k + 1)/s \rfloor$, giving
\begin{equation}\label{eq:feature_dim}
d = h_{\text{pool}} \times w_{\text{pool}} \times n_{\text{filters}}.
\end{equation}
Because the QUBO matrix has size $(d{+}1)K \times (d{+}1)K$, every architectural choice (number of filters, kernel size, and pooling stride) directly controls the number of binary variables and quadratic couplers. For example, with the architecture used in our experiments ($8\times 8$ input, $3\times 3$ kernel, $2\times 2$ pooling, 2~filters), $d = 3 \times 3 \times 2 = 18$, and each per-class QUBO has $19K$ binary variables. Doubling the filter count to 4 would increase $d$ to~36 and the QUBO size to $37K$, roughly quadrupling the number of quadratic terms.

Note that feature extraction is performed entirely on classical hardware. The frozen convolutional layers execute a standard forward pass (convolution, ReLU, pooling) to produce the feature matrix~$\mathbf{X}$. The quantum annealer (or simulated annealer) is invoked only for solving the QUBO that optimizes the FC layer weights. While the filter \emph{values} (i.e., the specific learned patterns such as edge or texture detectors) influence the quality of the extracted features and therefore the entries of the Gram matrix~$\mathbf{G}$, they do not affect the QUBO structure or size; only the architectural hyperparameters (kernel size, number of filters, pooling stride) determine~$d$ and hence the QUBO dimensions.

\subsection{Algorithm}\label{sec:algorithm}

\begin{figure}
\centering
\includegraphics{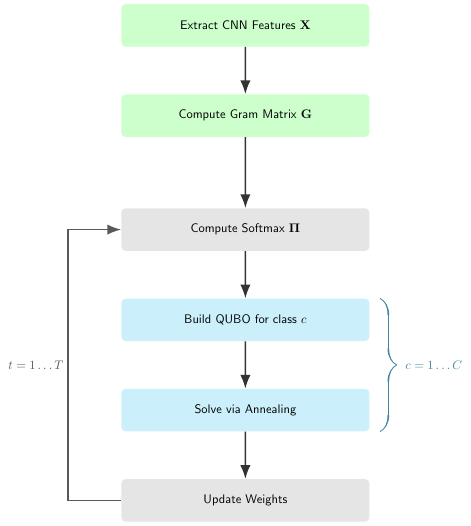}
\caption{Training pipeline matching Algorithm~\ref{alg:layerwise_gram}. Green blocks: one-time initialization (frozen CNN features and Gram matrix). Gray blocks: iterative phases~1 and~3. Cyan blocks: Phase~2 inner loop over $C$ classes. Outer loop repeats $T$ iterations.}
\label{fig:flowchart}
\end{figure}

The training procedure operates in four phases per iteration, as shown in Algorithm~\ref{alg:layerwise_gram}. Following the Extreme Learning Machine paradigm~\cite{park2019convolutional}, convolutional features are randomly initialized and frozen and only the FC layer is optimized, either via gradient descent~\cite{ruder2016overview} (classical baseline) or QUBO formulation (proposed method). After ReLU activation and max pooling, features are flattened and augmented with a bias column $\mathbf{X}_{\text{aug}} = [\mathbf{X}, \mathbf{1}_N] \in \mathbb{R}^{N \times (d+1)}$, where $d = h_{\text{pool}} \times w_{\text{pool}} \times n_{\text{filters}}$. The Gram matrix $\mathbf{G} = \frac{1}{N}\mathbf{X}_{\text{aug}}^T\mathbf{X}_{\text{aug}}$ is computed once and shared across all per-output QUBOs and iterations. At each iteration, softmax probabilities are computed, and for each output class $c$, an independent QUBO is formulated from the residual $\mathbf{r}_c = \mathbf{y}_c - \boldsymbol{\pi}_c$ and solved via simulated annealing (or quantum annealing). The binary solution decodes to a continuous weight update that is applied directly. Figure~\ref{fig:flowchart} illustrates the training pipeline, and Algorithm~\ref{alg:layerwise_gram} presents the complete procedure.

\begin{algorithm}
\caption{Layer-wise Gram Matrix QUBO Training}
\label{alg:layerwise_gram}
\begin{algorithmic}[1]
\Require Training data $\{(\mathbf{x}_n, \mathbf{y}_n)\}_{n=1}^N$, $C$ classes
\Require Frozen conv $\mathbf{W}_{\text{conv}}$, FC weights $\mathbf{W}_{\text{fc}} \in \mathbb{R}^{d \times C}$, biases $\mathbf{b}_{\text{fc}} \in \mathbb{R}^C$
\Require $\lambda$, precision $\mathbf{p} \in \mathbb{R}^K$, max update $\delta_{\max}$, iterations $T$
\Ensure Trained FC layer parameters
\State Initialize $\mathbf{W}_{\text{fc}}$, $\mathbf{b}_{\text{fc}}$ randomly
\State $\mathbf{W}_{\text{aug}} \gets [\mathbf{W}_{\text{fc}}; \mathbf{b}_{\text{fc}}^T]$ \Comment{Augmented weights}
\State $\mathbf{X} \gets \text{CNN\_features}(\mathbf{W}_{\text{conv}}, \{\mathbf{x}_n\})$ \Comment{Frozen features}
\State $\mathbf{X}_{\text{aug}} \gets [\mathbf{X}, \mathbf{1}_N]$ \Comment{Bias augmentation}
\State $\mathbf{G} \gets \frac{1}{N}\mathbf{X}_{\text{aug}}^T \mathbf{X}_{\text{aug}}$ \Comment{Gram matrix}
\For{$t = 1$ to $T$}
    \State $\boldsymbol{\Pi} \gets \text{softmax}(\mathbf{X}_{\text{aug}} \mathbf{W}_{\text{aug}})$
    \For{$c = 1$ to $C$}
        \State $\mathbf{r}_c \gets \mathbf{y}_c - \boldsymbol{\pi}_c$ \Comment{Residual}
        \State $\mathbf{g}_c \gets -\frac{1}{N}\mathbf{X}_{\text{aug}}^T \mathbf{r}_c + \lambda[\mathbf{w}_{c}; 0]$ \Comment{Gradient}
        \State $\mathbf{Q}_c \gets 4\mathcal{P}^T \mathbf{G}_\lambda \mathcal{P}$ \Comment{QUBO quadratic}
        \State $\mathbf{q}_c \gets 4\mathcal{P}^T(\mathbf{g}_c - \mathbf{G}_\lambda \delta_{\max}\mathbf{1}_{d+1})$ \Comment{QUBO linear}
        \State $\mathbf{b}_c^* \gets \text{Anneal}(\mathbf{Q}_c, \mathbf{q}_c)$ \Comment{Solve}
        \State $\mathbf{u}_c \gets 2\mathcal{P}\mathbf{b}_c^* - \delta_{\max}\mathbf{1}_{d+1}$ \Comment{Decode}
    \EndFor
    \State $\mathbf{W}_{\text{aug}} \gets \mathbf{W}_{\text{aug}} + \mathbf{U}_{\text{aug}}$
\EndFor
\State \Return $(\mathbf{W}_{\text{fc}}, \mathbf{b}_{\text{fc}})$
\end{algorithmic}
\end{algorithm}

\section{Experimental Results}\label{sec:results}

\begin{figure}
\centering
\includegraphics{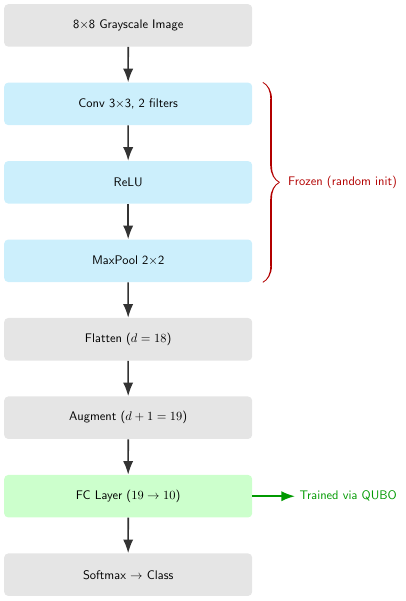}
\caption{CNN architecture used in all experiments. Convolutional layers are randomly initialized and frozen; only the fully connected (FC) classifier head is trained via iterative QUBO solves.}
\label{fig:cnn_architecture}
\end{figure}

This section presents experimental results in two parts: 1) detailed analysis on the \texttt{scikit-learn} digits dataset~\cite{pedregosa2011scikit} (Section~\ref{sec:sklearn_results}), which demonstrates convergence behavior and sample predictions, 2) followed by a multi-dataset benchmark (Section~\ref{sec:benchmark}) that validates generalization across five standard image classification tasks. All QUBO instances are solved using the \texttt{D-Wave Ocean} SDK~\cite{dwave2023ocean} with the simulated annealing sampler. All experiments use simulated annealing (SA), a classical optimization algorithm~\cite{kirkpatrick1983optimization}, rather than actual quantum hardware. SA mimics thermal annealing by probabilistically accepting worse solutions at high temperatures and gradually cooling to converge. While SA can effectively solve QUBO problems, it does not exhibit quantum effects. Our results establish baseline solution quality that quantum hardware would need to match or exceed. Figure~\ref{fig:cnn_architecture} illustrates the CNN architecture used across all experiments.

\subsection{\texttt{sklearn} Digits Dataset}\label{sec:sklearn_results}

\begin{table}
\centering
\caption{Experimental Settings (\texttt{sklearn} Digits Dataset)}
\label{tab:settings}
\begin{tabular}{ll}
\toprule
\textbf{Parameter} & \textbf{Value} \\
\midrule
Dataset & \texttt{sklearn} digits (8$\times$8 images) \\
Training samples & 1000 \\
Test samples & 540 \\
Number of classes & 10 \\
CNN architecture & 1 conv layer ($3\times3$ kernel), $2\times2$ pooling, FC output \\
CNN filters & 2 \\
Total parameters & 210 \\
Bit precisions & 5, 10, 15, 20 \\
Iterations & 1000 \\
Delta range & $[-0.5, 0.5]$ \\
Regularization ($\lambda$) & 0.001 \\
SA \texttt{num\_reads} & 1 (scaled with bits) \\
SA \texttt{num\_sweeps} & 1000 \\
SA $\beta$ range & $(0.01, 3)$ \\
\bottomrule
\end{tabular}
\end{table}

The SA solver has two key parameters that control the exploration-exploitation trade-off. \texttt{num\_sweeps} controls how long each SA run explores: each sweep performs one update attempt per binary variable. \texttt{num\_reads} controls how many independent SA runs are performed from different random initializations. With iterative QUBO training, we use $\texttt{num\_reads}=1$ because each iteration builds on the previous weights, providing natural error correction over time. 
Table~\ref{tab:settings} summarizes the experimental settings for the \texttt{sklearn} digits dataset.

The QUBO dimensions depend on feature dimension $d$, number of classes $C$, and bit precision $K$. For each per-output QUBO, the number of logical variables is $n = (d+1) \times K$, and the number of quadratic terms is $\frac{n(n-1)}{2}$. For our setup with $d = 18$ and $C = 10$ classes, Table~\ref{tab:qubo_size} shows the QUBO sizes for each bit precision.

\begin{table}
\centering
\caption{Per-Output QUBO Size by Bit Precision for all datasets ($d=18$, one of $C=10$ independent QUBOs)}
\label{tab:qubo_size}
\begin{tabular}{crr}
\toprule
$K$ (bits) & Logical Vars $n=(d{+}1)K$ & Quadratic Terms $\frac{n(n-1)}{2}$ \\
\midrule
5 & 95 & 4,465 \\
10 & 190 & 17,955 \\
15 & 285 & 40,470 \\
20 & 380 & 72,010 \\
\bottomrule
\end{tabular}
\end{table}

Since we solve $C$ independent QUBOs sequentially, hardware feasibility is determined by the largest single per-output QUBO, which is 380 qubits for 20-bit precision with $d=18$. All configurations remain well within D-Wave Advantage's 5,640 physical qubit limit. The test accuracy increases monotonically with bit precision, culminating in 81.5\% for 20-bit precision, as shown in Table~\ref{tab:results}, which is greater than the classical FC baseline 79.8\% accuracy. However it is important to note that training time is longer than classical training time.

\begin{table}
\centering
\caption{Layer-wise Gram QUBO Results on \texttt{sklearn} Digits (1000 iterations)}
\label{tab:results}
\begin{tabular}{lcccc}
\toprule
\textbf{Method} & \textbf{Train Acc} & \textbf{Test Acc} & \textbf{Time (s)} & \textbf{Final Loss} \\
\midrule
Classical FC (SGD) & 82.2\% & 79.8\% & 22.7 & 0.872 \\
\midrule
QUBO 5-bit & 33.9\% & 33\% & 225.6 & 2.704 \\
QUBO 10-bit & 77.7\% & 77.4\% & 767.3 & 0.758 \\
QUBO 15-bit & 83.9\% & 80\% & 2605.4 & 0.677 \\
QUBO 20-bit & 84.9\% & \textbf{81.5\%} & 9712.6 & 0.671 \\
\bottomrule
\end{tabular}
\end{table}

\begin{figure}[h]
\centering
\begin{subfigure}[b]{0.48\textwidth}
\includegraphics[width=\textwidth]{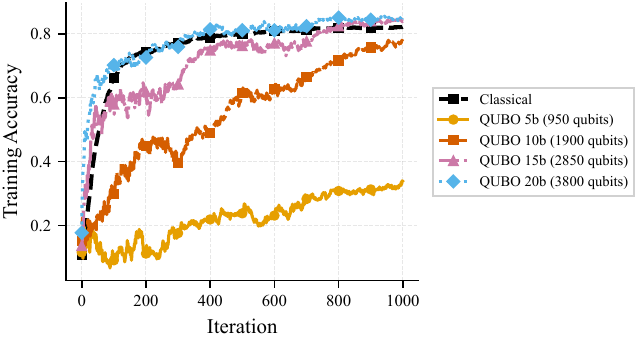}
\caption{Training Accuracy}
\end{subfigure}
\hfill
\begin{subfigure}[b]{0.48\textwidth}
\includegraphics[width=\textwidth]{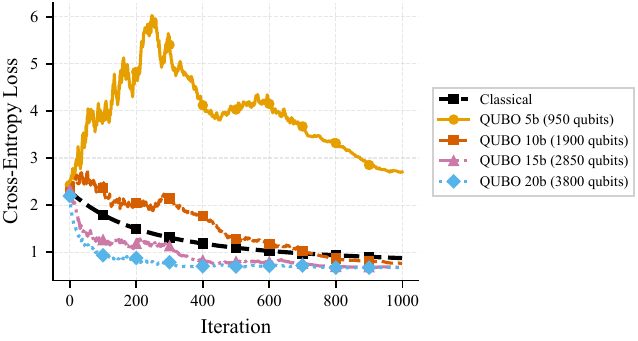}
\caption{Training Loss}
\end{subfigure}
\caption{Convergence curves for Classical FC (SGD) and QUBO methods with different bit precisions over 1000 iterations.}
\label{fig:convergence}
\end{figure}

QUBO methods exhibit stepwise convergence with fluctuations, since updates are applied unconditionally and some iterations increase the loss before subsequent iterations correct them. The 20-bit configuration converges fastest among QUBO methods, exceeding classical accuracy by iteration 500 (81.2\% vs 79.8\%), as shown in Fig.~\ref{fig:convergence}. Both 15-bit and 20-bit achieve lower final loss (0.677, 0.671) than classical (0.872), indicating that the QUBO surrogate with sufficient precision can find better solutions than SGD in this frozen-feature setting. The 5-bit configuration fails to converge meaningfully, with loss remaining high ($>$2.7) throughout training.

\begin{figure}[H]
\centering
\includegraphics[width=0.9\textwidth]{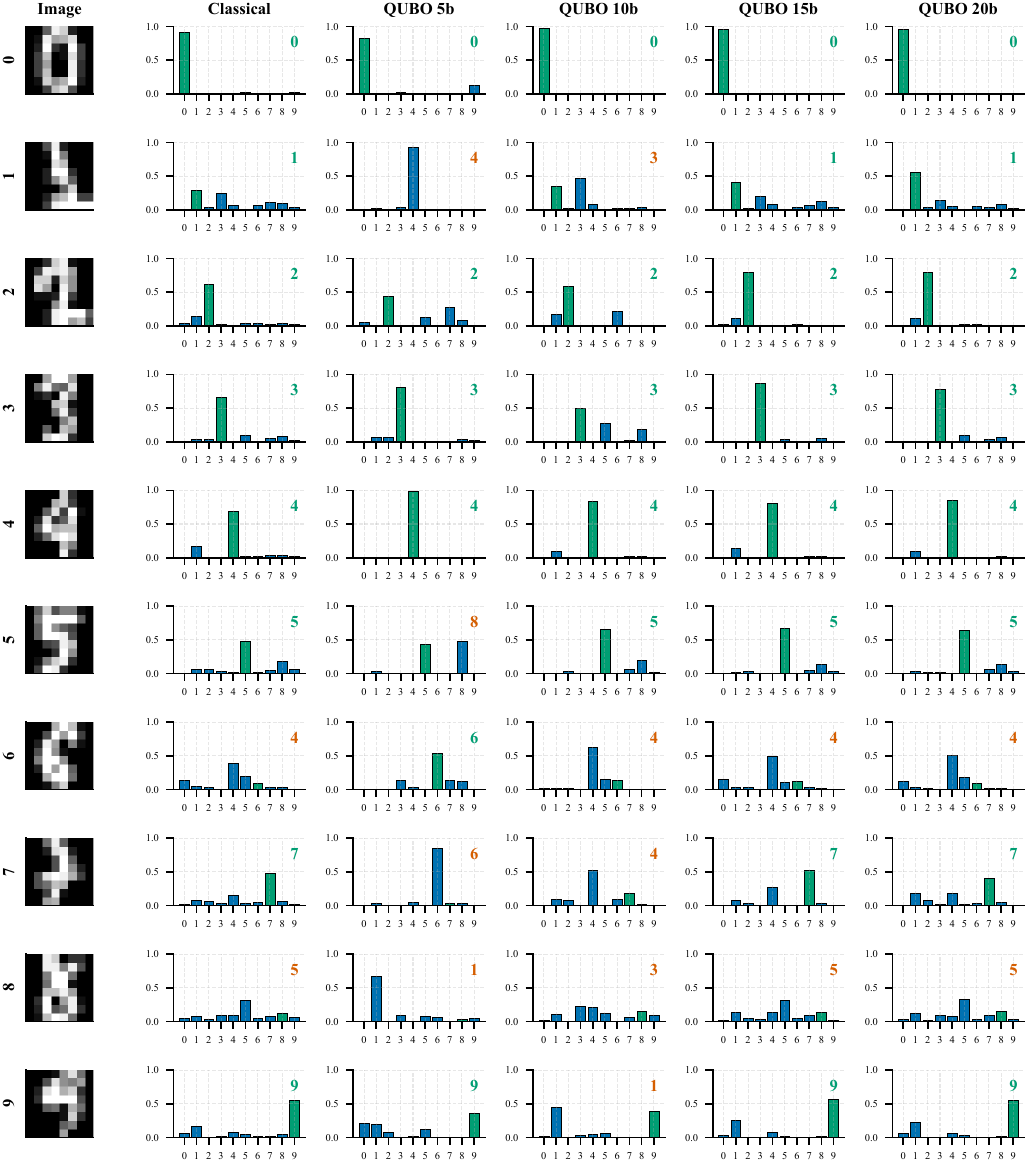}
\caption{Sample predictions from Classical FC (SGD) and QUBO models with different bit precisions. Each row shows predictions for one test digit, with columns showing probability distributions from each model.}
\label{fig:predictions}
\end{figure}

Digits 0, 2, 3, and 4 are correctly classified by all models, while Digit 8 is misclassified by all models, and the remaining digits show mixed results, as seen in Fig.~\ref{fig:predictions} and Table~\ref{tab:sample_predictions}. The universal failure on Digit 8 suggests that the $8\times 8$ representation, rather than the optimizer, is the bottleneck: at this resolution, Digit 8 loses the distinguishing loop structure that separates it from visually similar digits. 

\begin{table}[h]
\centering
\caption{Sample Prediction Analysis: Probability of Correct Class. T = correct prediction (True), F = incorrect prediction (False).}
\label{tab:sample_predictions}
\scriptsize
\setlength{\tabcolsep}{3pt}
\begin{tabular}{ccccccl}
\toprule
\textbf{True Digit} & \textbf{Classical} & \textbf{5-bit} & \textbf{10-bit} & \textbf{15-bit} & \textbf{20-bit} & \textbf{Result} \\
\midrule
0 & 91.6\% (T) & 83.4\% (T) & 98.1\% (T) & 96\% (T) & 96.7\% (T) & All correct \\
1 & 29.8\% (T) & 1.8\% (F) & 34.6\% (F) & 41.6\% (T) & 55.3\% (T) & Mixed \\
2 & 61.6\% (T) & 43.8\% (T) & 58.3\% (T) & 79.7\% (T) & 78.9\% (T) & All correct \\
3 & 66.3\% (T) & 80.4\% (T) & 49.9\% (T) & 86.2\% (T) & 77.7\% (T) & All correct \\
4 & 68.3\% (T) & 98.1\% (T) & 84.2\% (T) & 80.3\% (T) & 85.3\% (T) & All correct \\
5 & 48\% (T) & 44.2\% (F) & 66.2\% (T) & 67.3\% (T) & 64.2\% (T) & Mixed \\
6 & 9.6\% (F) & 54.6\% (T) & 62.9\% (F) & 48.9\% (F) & 50.6\% (F) & Mixed \\
7 & 47.6\% (T) & 85.5\% (F) & 52.9\% (F) & 53.1\% (T) & 40\% (T) & Mixed \\
8 & 32.3\% (F) & 67.6\% (F) & 22.5\% (F) & 31.3\% (F) & 32.8\% (F) & All wrong \\
9 & 55.6\% (T) & 35.7\% (T) & 38.9\% (F) & 57\% (T) & 56\% (T) & Mixed \\
\bottomrule
\end{tabular}
\end{table}

\subsection{Multi-Dataset Benchmark}\label{sec:benchmark}
To validate generalization beyond the \texttt{sklearn} digits dataset, we evaluate on five standard image classification benchmarks: MNIST, Fashion-MNIST, CIFAR-10, EMNIST (letters A--J), and KMNIST. All datasets are downsampled to 8$\times$8 grayscale images. We report mean $\pm$ standard deviation across 5 random seeds. For each dataset, we subsample 1000 training and 500 test samples with approximately uniform class coverage (100 and 50 samples per class respectively for $C=10$). The frozen convolutional filters are reinitialized per seed to capture realistic random-feature variability. Table~\ref{tab:benchmark_settings} summarizes the benchmark configuration.

\begin{table}[h]
\centering
\caption{Benchmark Experimental Settings}
\label{tab:benchmark_settings}
\begin{tabular}{ll}
\toprule
\textbf{Parameter} & \textbf{Value} \\
\midrule
Datasets & MNIST, Fashion-MNIST, CIFAR-10, EMNIST, KMNIST \\
Image resolution & 8$\times$8 grayscale \\
Training samples & 1000 \\
Test samples & 500 \\
Number of classes & 10 \\
Random seeds & 5 (42, 43, 44, 45, 46) \\
CNN architecture & 1 conv layer ($3\times3$ kernel), $2\times2$ pooling, FC output \\
CNN filters & 2 \\
Iterations & 1000 \\
Bit precisions & 5, 10, 15, 20 \\
Delta range & $[-0.5, 0.5]$ \\
Regularization ($\lambda$) & 0.001 \\
SA \texttt{num\_sweeps} & 100 \\
SA \texttt{num\_reads} & 1 \\
SA $\beta$ range & $(0.01, 3)$ \\
\bottomrule
\end{tabular}
\end{table}

The benchmark experiments use $\texttt{num\_sweeps}=100$ (vs.\ 1000 in Section~\ref{sec:sklearn_results}), reducing per-QUBO solve time by approximately $10\times$. This choice reflects a trade-off: with 5 seeds $\times$ 5 methods $\times$ 5 datasets $=$ 125 training runs, computational cost becomes significant. Empirically, the reduced sweep count still produces competitive results because iterative training provides natural error correction, as occasional suboptimal SA solutions are corrected by subsequent iterations.

While the precision study in Section~\ref{sec:sklearn_results} focused on accuracy and loss to isolate the effect of bit precision, the multi-dataset benchmark uses six complementary metrics to provide a more comprehensive evaluation across diverse datasets. For multiclass classification with $C$ classes, metrics are computed in a one-vs-rest (OvR) manner and macro-averaged across classes. Let TP$_c$, TN$_c$, FP$_c$, FN$_c$ denote the true positives, true negatives, false positives, and false negatives for class $c$. Table~\ref{tab:metrics_definition} summarizes the metrics.

\begin{table}[h]
\centering
\caption{Evaluation Metrics Summary. For all metrics, higher values indicate better performance. The Chance Level column shows the expected value for a random classifier with $C=10$ equiprobable classes.}
\label{tab:metrics_definition}
\scriptsize
\setlength{\tabcolsep}{3pt}
\begin{tabular}{lccc}
\toprule
\textbf{Metric} & \textbf{Formula} & \textbf{Range} & \textbf{Chance Level ($C{=}10$)} \\
\midrule
Accuracy & $\frac{\text{TP}+\text{TN}}{\text{TP}+\text{TN}+\text{FP}+\text{FN}}$ & $[0, 1]$ & 10\% \\
Precision & $\frac{\text{TP}}{\text{TP}+\text{FP}}$ & $[0, 1]$ & 10\% \\
Recall & $\frac{\text{TP}}{\text{TP}+\text{FN}}$ & $[0, 1]$ & 10\% \\
F1 Score & $\frac{2 \cdot \text{Prec} \cdot \text{Rec}}{\text{Prec}+\text{Rec}}$ & $[0, 1]$ & 10\% \\
Cohen's $\kappa$ & $\frac{p_o - p_e}{1 - p_e}$ & $[-1, 1]$ & 0\% \\
MCC & $\frac{\text{TP}\cdot\text{TN}-\text{FP}\cdot\text{FN}}{\sqrt{(\text{TP}+\text{FP})(\text{TP}+\text{FN})(\text{TN}+\text{FP})(\text{TN}+\text{FN})}}$ & $[-1, 1]$ & 0\% \\
\bottomrule
\end{tabular}
\end{table}

Accuracy measures the fraction of samples correctly classified. Precision measures the fraction of predicted positives that are true positives, while recall measures the fraction of actual positives correctly identified. Both are macro-averaged across classes. F1 is the harmonic mean of precision and recall, penalizing extreme imbalance. Cohen's Kappa~\cite{cohen1960coefficient} measures agreement beyond chance, where $p_o$ is observed agreement and $p_e$ is expected agreement under random assignment. $\kappa < 0.2$ indicates slight agreement, $0.4$--$0.6$ moderate, and $>0.8$ near-perfect. The Matthews Correlation Coefficient (MCC)~\cite{matthews1975comparison} is a correlation coefficient between observed and predicted classifications, robust to class imbalance. Accuracy alone can be misleading for imbalanced data. Kappa and MCC correct for chance agreement, while precision, recall, and F1~\cite{van1979information} reveal class-specific behavior.

Table~\ref{tab:benchmark_consolidated} presents a comparison across all five datasets (mean $\pm$ std over 5 seeds). QUBO 20-bit achieves the strongest results on MNIST (+3.1\% test accuracy over classical) and Fashion-MNIST (+1.3\%), with improvements that exceed one standard deviation. On EMNIST, QUBO 20-bit matches the classical baseline ($\pm$0\%). CIFAR-10 and KMNIST show slightly lower QUBO accuracy, which we attribute to the difficulty of these datasets at $8\times 8$ resolution combined with only two convolutional filters.

\begin{table}[h]
\centering
\caption{Consolidated Benchmark Results (\%, mean $\pm$ std over 5 seeds). Best per dataset in bold.}
\label{tab:benchmark_consolidated}
\tiny
\setlength{\tabcolsep}{2pt}
\begin{tabular}{ll|ccccccc}
\toprule
\textbf{Dataset} & \textbf{Method} & \textbf{Train Acc} & \textbf{Test Acc} & \textbf{Precision} & \textbf{Recall} & \textbf{F1} & \textbf{Kappa} & \textbf{MCC} \\
\midrule
\multirow{5}{*}{MNIST}
& Classical & $81.6 \pm 0.5$ & $78.2 \pm 1.9$ & $78.7 \pm 1.5$ & $78.1 \pm 1.7$ & $77.8 \pm 1.8$ & $75.8 \pm 1.8$ & $75.9 \pm 1.8$ \\
& QUBO 5b & $37.2 \pm 3.9$ & $36.5 \pm 4.2$ & $35.9 \pm 3.9$ & $36.4 \pm 3.7$ & $35.5 \pm 3.6$ & $29.5 \pm 4.1$ & $29.6 \pm 4.1$ \\
& QUBO 10b & $76.5 \pm 1.8$ & $74.4 \pm 2.9$ & $74.7 \pm 2.4$ & $74.2 \pm 2.8$ & $74 \pm 2.7$ & $71.6 \pm 2.9$ & $71.6 \pm 2.9$ \\
& QUBO 15b & $83.7 \pm 0.5$ & $80.1 \pm 1.9$ & $80.3 \pm 1.7$ & $80 \pm 1.7$ & $79.8 \pm 1.9$ & $77.9 \pm 1.8$ & $78 \pm 1.8$ \\
& QUBO 20b & $\mathbf{84 \pm 1}$ & $\mathbf{81.3 \pm 1.4}$ & $\mathbf{81.6 \pm 1.2}$ & $\mathbf{81.1 \pm 1.2}$ & $\mathbf{80.9 \pm 1.3}$ & $\mathbf{79.2 \pm 1.4}$ & $\mathbf{79.3 \pm 1.4}$ \\
\midrule
\multirow{5}{*}{Fashion}
& Classical & $54.5 \pm 0.9$ & $52.1 \pm 1.1$ & $52.6 \pm 2.3$ & $51.8 \pm 0.4$ & $50 \pm 1$ & $46.7 \pm 1$ & $47.1 \pm 1$ \\
& QUBO 5b & $33.1 \pm 2.6$ & $32.6 \pm 2.3$ & $32.1 \pm 2.7$ & $32.5 \pm 2.3$ & $31.4 \pm 2.6$ & $25.1 \pm 2.3$ & $25.2 \pm 2.3$ \\
& QUBO 10b & $52 \pm 1.6$ & $46.6 \pm 1.9$ & $46.7 \pm 1.6$ & $46.4 \pm 1.5$ & $45.2 \pm 1.4$ & $40.6 \pm 1.9$ & $40.9 \pm 1.9$ \\
& QUBO 15b & $56.6 \pm 1.5$ & $53.1 \pm 1.4$ & $52.7 \pm 1.2$ & $52.8 \pm 1.1$ & $51.5 \pm 1.6$ & $47.8 \pm 1.4$ & $48.1 \pm 1.2$ \\
& QUBO 20b & $\mathbf{56.8 \pm 1.2}$ & $\mathbf{53.4 \pm 1.7}$ & $\mathbf{52.9 \pm 1.1}$ & $\mathbf{53.3 \pm 0.8}$ & $\mathbf{52.1 \pm 1.3}$ & $\mathbf{48.2 \pm 1.6}$ & $\mathbf{48.4 \pm 1.5}$ \\
\midrule
\multirow{5}{*}{CIFAR-10}
& Classical & $23.5 \pm 1$ & $\mathbf{21.6 \pm 2.1}$ & $21.2 \pm 2.4$ & $\mathbf{21.6 \pm 1.6}$ & $19 \pm 2.5$ & $\mathbf{12.8 \pm 1.9}$ & $\mathbf{13.2 \pm 1.8}$ \\
& QUBO 5b & $14.8 \pm 1.4$ & $14.4 \pm 2.9$ & $14.4 \pm 3.1$ & $14.4 \pm 2.6$ & $13.5 \pm 2.3$ & $4.9 \pm 2.9$ & $5 \pm 3$ \\
& QUBO 10b & $22.7 \pm 1.3$ & $19.7 \pm 3.1$ & $19.3 \pm 3.4$ & $19.6 \pm 2.5$ & $18.3 \pm 3.2$ & $10.7 \pm 2.9$ & $10.9 \pm 2.9$ \\
& QUBO 15b & $24.2 \pm 1$ & $21.4 \pm 2$ & $20.9 \pm 2.4$ & $21.3 \pm 1.6$ & $19.3 \pm 2.2$ & $12.7 \pm 1.9$ & $12.9 \pm 1.9$ \\
& QUBO 20b & $\mathbf{24.9 \pm 1.2}$ & $21.4 \pm 1.4$ & $\mathbf{21.6 \pm 2.6}$ & $21.3 \pm 1$ & $\mathbf{19.4 \pm 1.6}$ & $12.6 \pm 1.2$ & $12.9 \pm 1.1$ \\
\midrule
\multirow{5}{*}{EMNIST}
& Classical & $52.2 \pm 1.7$ & $\mathbf{47.3 \pm 1.2}$ & $\mathbf{46.5 \pm 1.5}$ & $46.7 \pm 0.9$ & $45.4 \pm 0.8$ & $41.3 \pm 1.2$ & $41.5 \pm 1.2$ \\
& QUBO 5b & $19.4 \pm 5.5$ & $19.2 \pm 6.6$ & $19.4 \pm 5.9$ & $19.2 \pm 5.9$ & $18.9 \pm 5.8$ & $10.2 \pm 6.6$ & $10.2 \pm 6.6$ \\
& QUBO 10b & $43.3 \pm 1.1$ & $39.4 \pm 1.3$ & $38.9 \pm 0.7$ & $38.9 \pm 0.9$ & $38.1 \pm 0.5$ & $32.5 \pm 1.2$ & $32.7 \pm 1.3$ \\
& QUBO 15b & $51.4 \pm 2$ & $46.7 \pm 2.8$ & $45.6 \pm 2.3$ & $46 \pm 2.3$ & $45 \pm 2.3$ & $40.6 \pm 2.7$ & $40.8 \pm 2.7$ \\
& QUBO 20b & $\mathbf{52.8 \pm 1.9}$ & $\mathbf{47.3 \pm 2.3}$ & $\mathbf{46.5 \pm 1.9}$ & $\mathbf{46.8 \pm 1.8}$ & $\mathbf{45.6 \pm 1.6}$ & $\mathbf{41.4 \pm 2.3}$ & $\mathbf{41.6 \pm 2.3}$ \\
\midrule
\multirow{5}{*}{KMNIST}
& Classical & $\mathbf{49.8 \pm 1.6}$ & $\mathbf{44.7 \pm 2.6}$ & $\mathbf{43.4 \pm 2}$ & $\mathbf{44.2 \pm 2.3}$ & $\mathbf{42.6 \pm 2.2}$ & $\mathbf{38.5 \pm 2.6}$ & $\mathbf{38.7 \pm 2.6}$ \\
& QUBO 5b & $19.6 \pm 1$ & $18.4 \pm 1.7$ & $18.7 \pm 2.2$ & $18.3 \pm 1.4$ & $18.1 \pm 1.5$ & $9.3 \pm 1.7$ & $9.4 \pm 1.7$ \\
& QUBO 10b & $41.1 \pm 1.3$ & $35.6 \pm 3$ & $35.2 \pm 2.1$ & $35.1 \pm 2.4$ & $34.5 \pm 2.4$ & $28.4 \pm 2.9$ & $28.6 \pm 2.9$ \\
& QUBO 15b & $48.1 \pm 1.4$ & $43 \pm 2.9$ & $41.8 \pm 2.5$ & $42.4 \pm 2.5$ & $41.3 \pm 2.6$ & $36.6 \pm 2.9$ & $36.7 \pm 2.9$ \\
& QUBO 20b & $49.6 \pm 1.1$ & $43.4 \pm 2.7$ & $42 \pm 2.1$ & $42.8 \pm 2.1$ & $41.6 \pm 2$ & $37 \pm 2.6$ & $37.2 \pm 2.6$ \\
\bottomrule
\end{tabular}
\end{table}

Table~\ref{tab:recall_all} presents per-class recall (sensitivity) for all datasets, computed from confusion matrices. Recall for class $c$ measures the fraction of true class $c$ samples correctly identified: $\text{Recall}_c = \text{TP}_c / (\text{TP}_c + \text{FN}_c)$. We report recall rather than one-vs-rest accuracy because OvR accuracy is dominated by true negatives in balanced multiclass settings. Values are mean $\pm$ std (\%) over 5 seeds.

\begin{table}[h]
\centering
\caption{Per-Class Recall Across All Datasets (\%, mean $\pm$ std over 5 seeds). All datasets have $C=10$ classes: MNIST/Fashion/CIFAR-10 use digits/items/objects 0--9; EMNIST uses letters A--J; KMNIST uses 10 hiragana characters.}
\label{tab:recall_all}
\tiny
\setlength{\tabcolsep}{2pt}
\begin{tabular}{ll|cccccccccc}
\toprule
\textbf{Dataset} & \textbf{Method} & \textbf{0} & \textbf{1} & \textbf{2} & \textbf{3} & \textbf{4} & \textbf{5} & \textbf{6} & \textbf{7} & \textbf{8} & \textbf{9} \\
\midrule
\multirow{5}{*}{MNIST}
& Classical & \textbf{98}$\pm$1 & 71$\pm$7 & 91$\pm$4 & 80$\pm$6 & \textbf{88}$\pm$3 & 65$\pm$7 & 88$\pm$1 & 73$\pm$5 & 49$\pm$6 & \textbf{78}$\pm$5 \\
& QUBO 5b & 51$\pm$26 & 22$\pm$11 & 31$\pm$20 & 37$\pm$14 & 38$\pm$10 & 33$\pm$11 & 48$\pm$16 & 47$\pm$13 & 22$\pm$14 & 35$\pm$20 \\
& QUBO 10b & 95$\pm$5 & 61$\pm$3 & 75$\pm$12 & 75$\pm$3 & 85$\pm$3 & 68$\pm$5 & \textbf{90}$\pm$2 & 71$\pm$5 & 50$\pm$7 & 73$\pm$9 \\
& QUBO 15b & \textbf{98}$\pm$1 & 77$\pm$7 & \textbf{92}$\pm$4 & 83$\pm$3 & 87$\pm$5 & 70$\pm$6 & 88$\pm$2 & 73$\pm$5 & \textbf{57}$\pm$8 & 74$\pm$7 \\
& QUBO 20b & \textbf{98}$\pm$2 & \textbf{78}$\pm$7 & \textbf{92}$\pm$3 & \textbf{84}$\pm$4 & \textbf{88}$\pm$3 & \textbf{73}$\pm$4 & \textbf{90}$\pm$2 & \textbf{76}$\pm$5 & 53$\pm$8 & \textbf{78}$\pm$4 \\
\midrule
\multirow{5}{*}{Fashion}
& Classical & \textbf{64}$\pm$6 & 71$\pm$12 & 36$\pm$12 & 49$\pm$15 & 27$\pm$15 & \textbf{70}$\pm$10 & 20$\pm$8 & 64$\pm$20 & 53$\pm$11 & 64$\pm$12 \\
& QUBO 5b & 31$\pm$6 & 64$\pm$7 & 20$\pm$6 & 28$\pm$12 & 16$\pm$6 & 40$\pm$16 & 18$\pm$8 & 45$\pm$14 & 24$\pm$11 & 37$\pm$8 \\
& QUBO 10b & 54$\pm$7 & 55$\pm$16 & 28$\pm$7 & 49$\pm$12 & \textbf{29}$\pm$12 & 60$\pm$14 & 20$\pm$10 & 58$\pm$20 & 46$\pm$13 & 65$\pm$6 \\
& QUBO 15b & \textbf{64}$\pm$5 & 67$\pm$16 & \textbf{42}$\pm$11 & \textbf{50}$\pm$16 & 28$\pm$10 & 66$\pm$11 & 22$\pm$6 & 65$\pm$18 & 56$\pm$8 & \textbf{69}$\pm$9 \\
& QUBO 20b & 63$\pm$7 & \textbf{72}$\pm$12 & \textbf{42}$\pm$5 & 47$\pm$11 & 27$\pm$8 & 66$\pm$12 & \textbf{24}$\pm$8 & \textbf{67}$\pm$15 & \textbf{58}$\pm$8 & 66$\pm$7 \\
\midrule
\multirow{5}{*}{CIFAR-10}
& Classical & \textbf{32}$\pm$7 & \textbf{21}$\pm$16 & 13$\pm$8 & 8$\pm$8 & \textbf{36}$\pm$20 & 8$\pm$7 & 13$\pm$11 & \textbf{36}$\pm$11 & 27$\pm$5 & 21$\pm$10 \\
& QUBO 5b & 16$\pm$4 & 14$\pm$5 & \textbf{25}$\pm$15 & 9$\pm$4 & 15$\pm$14 & 7$\pm$4 & 9$\pm$5 & 17$\pm$7 & 21$\pm$5 & 11$\pm$5 \\
& QUBO 10b & 30$\pm$4 & 15$\pm$4 & 11$\pm$9 & \textbf{11}$\pm$8 & 34$\pm$20 & 9$\pm$4 & 16$\pm$9 & 20$\pm$9 & 26$\pm$6 & 23$\pm$9 \\
& QUBO 15b & 28$\pm$7 & 19$\pm$9 & 13$\pm$12 & 9$\pm$5 & 35$\pm$20 & \textbf{10}$\pm$7 & \textbf{19}$\pm$13 & 30$\pm$9 & \textbf{30}$\pm$8 & 20$\pm$8 \\
& QUBO 20b & 29$\pm$6 & 18$\pm$8 & 19$\pm$11 & 10$\pm$7 & 34$\pm$20 & 9$\pm$6 & 15$\pm$11 & 31$\pm$11 & 26$\pm$5 & \textbf{24}$\pm$9 \\
\midrule
\multirow{5}{*}{EMNIST}
& Classical & 36$\pm$6 & 43$\pm$8 & \textbf{67}$\pm$5 & \textbf{36}$\pm$8 & 26$\pm$7 & \textbf{74}$\pm$4 & 24$\pm$4 & 38$\pm$14 & 68$\pm$5 & 54$\pm$6 \\
& QUBO 5b & 12$\pm$7 & 27$\pm$8 & 23$\pm$13 & 15$\pm$10 & 14$\pm$8 & 29$\pm$14 & 17$\pm$9 & 17$\pm$11 & 22$\pm$20 & 17$\pm$6 \\
& QUBO 10b & 23$\pm$5 & 34$\pm$10 & 56$\pm$11 & 32$\pm$12 & 26$\pm$7 & 55$\pm$6 & 21$\pm$4 & 35$\pm$9 & 59$\pm$10 & 47$\pm$8 \\
& QUBO 15b & 33$\pm$10 & \textbf{44}$\pm$9 & 64$\pm$6 & 31$\pm$9 & \textbf{33}$\pm$5 & 67$\pm$8 & 26$\pm$3 & \textbf{39}$\pm$12 & 68$\pm$8 & \textbf{55}$\pm$8 \\
& QUBO 20b & \textbf{37}$\pm$3 & 42$\pm$8 & \textbf{67}$\pm$9 & 33$\pm$7 & 27$\pm$7 & 71$\pm$7 & \textbf{27}$\pm$4 & \textbf{39}$\pm$13 & \textbf{74}$\pm$6 & 52$\pm$5 \\
\midrule
\multirow{5}{*}{KMNIST}
& Classical & \textbf{73}$\pm$5 & \textbf{33}$\pm$8 & 13$\pm$5 & \textbf{66}$\pm$7 & \textbf{32}$\pm$6 & \textbf{56}$\pm$8 & \textbf{51}$\pm$6 & 39$\pm$11 & \textbf{31}$\pm$11 & 48$\pm$9 \\
& QUBO 5b & 26$\pm$9 & 18$\pm$5 & 12$\pm$3 & 21$\pm$11 & 18$\pm$8 & 18$\pm$9 & 29$\pm$11 & 14$\pm$9 & 12$\pm$3 & 15$\pm$3 \\
& QUBO 10b & 54$\pm$9 & 29$\pm$7 & \textbf{17}$\pm$10 & 54$\pm$7 & 28$\pm$5 & 49$\pm$16 & 30$\pm$10 & 31$\pm$9 & 22$\pm$6 & 38$\pm$11 \\
& QUBO 15b & 71$\pm$4 & 32$\pm$8 & \textbf{17}$\pm$6 & 62$\pm$7 & 29$\pm$6 & \textbf{56}$\pm$8 & 47$\pm$10 & 36$\pm$9 & \textbf{31}$\pm$9 & 43$\pm$10 \\
& QUBO 20b & 68$\pm$8 & \textbf{33}$\pm$9 & 15$\pm$5 & 65$\pm$8 & 30$\pm$9 & 55$\pm$9 & 46$\pm$8 & \textbf{40}$\pm$10 & 29$\pm$9 & \textbf{46}$\pm$8 \\
\bottomrule
\end{tabular}
\end{table}

The QUBO problem sizes are identical across datasets since all use the same CNN architecture (Table~\ref{tab:qubo_size}). Table~\ref{tab:execution_times} reports execution times (mean $\pm$ std over 5 seeds). Full class label mappings for each dataset are provided in the code repository.\footnote{\url{https://github.com/Mostafa-Atallah2020/qml-qubo}}


\begin{table}[h]
\centering
\caption{Execution Time by Dataset and Bit Precision (seconds, mean $\pm$ std over 5 seeds)}
\label{tab:execution_times}
\scriptsize
\setlength{\tabcolsep}{3pt}
\begin{tabular}{lcccc}
\toprule
\textbf{Dataset} & \textbf{5-bit} & \textbf{10-bit} & \textbf{15-bit} & \textbf{20-bit} \\
\midrule
MNIST & $160 \pm 5$ & $448 \pm 48$ & $1,063 \pm 127$ & $2,449 \pm 229$ \\
Fashion & $157 \pm 9$ & $464 \pm 50$ & $1,128 \pm 127$ & $2,433 \pm 241$ \\
CIFAR-10 & $157 \pm 8$ & $466 \pm 49$ & $1,118 \pm 119$ & $2,351 \pm 192$ \\
EMNIST & $157 \pm 7$ & $464 \pm 49$ & $1,125 \pm 121$ & $2,370 \pm 199$ \\
KMNIST & $148 \pm 17$ & $438 \pm 55$ & $1,014 \pm 122$ & $2,424 \pm 234$ \\
\bottomrule
\end{tabular}
\end{table}

The QUBOs are fully dense (density = 1) due to the Gram matrix structure. Each iteration solves $C$ independent QUBOs, and $T$ iterations yield $C \times T$ total QUBO solves. For our experiments with $C=10$ and $T=1000$, this gives 10,000 QUBO solves per training run. All configurations remain within D-Wave Advantage's 5,640 physical qubit limit, though dense connectivity requires embedding overhead. Execution time scales approximately as $O((d{+}1)^2 K^2)$ due to the fully dense QUBO structure, consistent with the observed $\sim$15$\times$ increase from 5-bit to 20-bit. Across 5 random seeds, QUBO 20b achieves higher test accuracy than classical on 5/5 seeds for MNIST, 4/5 for Fashion-MNIST, 3/5 for EMNIST, 2/5 for CIFAR-10, and 1/5 for KMNIST. On MNIST and Fashion-MNIST, the improvements exceed one standard deviation and are statistically significant under a one-sided paired $t$-test ($p < 0.05$), indicating consistent gains beyond random seed variation. Notably, the 15-bit formulation (285 variables, 40,470 couplers) is the highest precision that fits within the 40,484 physical couplers of the current D-Wave Advantage Pegasus topology, making it the most practical configuration for near-term quantum hardware deployment.

\section{Discussion and Conclusion}\label{sec:discussion}

We presented a layer-wise Gram-matrix QUBO formulation for iterative training of CNN classifier heads that addresses three fundamental challenges in applying QUBOs to neural network training: 1) the non-quadratic cross-entropy loss is replaced by a convex quadratic surrogate (Equation~\eqref{eq:surrogate_reg}), 2) the non-convex loss landscape is navigated through iterative updates rather than one-shot optimization, and 3) the problem size is reduced through per-output decomposition into $C$ independent QUBOs of $(d+1)K$ variables each. Across six datasets and six evaluation metrics, QUBO 20-bit matches or exceeds classical SGD on MNIST (+3.1\% accuracy, +3.4\% Kappa), Fashion-MNIST (+1.3\%, +1.5\%), and EMNIST ($\pm$0\%), while remaining competitive on CIFAR-10 ($-$0.2\%) and KMNIST ($-$1.3\%). A minimum bit precision of $K \geq 10$ is required for effective optimization: the 5-bit configuration fails across all datasets (14--37\% accuracy), while 10 bits and above produce competitive results. This threshold is consistent across all six datasets, confirming it is a property of the QUBO formulation rather than the data. The training trajectory converges monotonically in the aggregate: although individual iterations may temporarily increase the loss (approximately 43--46\% of updates increase the loss, computed as the fraction of iterations where the loss after the QUBO update exceeds the loss before it), the overall accuracy improves consistently for $K\geq 10$, with standard deviations of 1--3\% across seeds indicating stable optimization. All QUBO updates are applied unconditionally with step size $\alpha = 1$ (full update). We tested $\alpha \in \{0.25, 0.5, 1.0\}$ in preliminary experiments and found that $\alpha = 1$ provided the fastest convergence without instability, since the quadratic surrogate already constrains update magnitudes via $\delta_{\max}$ and the iterative procedure naturally corrects occasional harmful updates.

The formulation has several implications and limitations. Each per-output QUBO with $d=18$ and $K=20$ has 380 logical variables, within D-Wave Advantage's 5,640-qubit capacity, but the dense connectivity (72,010 couplers) exceeds the 40,484 physical couplers in the Pegasus topology~\cite{dwave2020pegasus}. Preliminary minor-embedding estimates on the Pegasus topology suggest chain lengths of $\sim$19--25 qubits per logical variable for a fully-connected $K_{380}$ graph, consuming roughly 7,000--9,500 physical qubits. The recently released D-Wave Advantage2 (Zephyr topology, degree 20) offers higher per-qubit connectivity but still cannot embed $K_{380}$. Practical QPU deployment would require reduced bit precision (e.g., $K=5$ gives $K_{95}$, embeddable on current hardware), sparser QUBO approximations, or next-generation hardware. QUBO training is 100--400$\times$ slower than classical SGD using simulated annealing. Implementing quantum annealing on quantum hardware may offer speedups through quantum tunneling on rugged energy landscapes~\cite{kadowaki1998quantum,farhi2001quantum}, though this remains a question for future work. All experiments use frozen, randomly initialized convolutional features following the Extreme Learning Machine paradigm~\cite{huang2006extreme,park2019convolutional}, training only the FC classifier head. On CIFAR-10, QUBO 20-bit achieves 21.4\% versus 21.6\% for the classical baseline (near parity), indicating that the low absolute accuracy reflects the $8\times 8$ downsampling bottleneck rather than optimizer failure. The per-output decomposition ignores softmax coupling across classes, and all results use simulated annealing rather than quantum hardware. 

Several directions could extend this work. Validating the formulation on quantum annealing hardware would establish whether QPU solutions match or exceed SA quality. Adaptive bit precision, starting with low $K$ for coarse exploration and increasing $K$ as training converges, could reduce total QUBO solve time while maintaining final accuracy. Extending the Gram-matrix surrogate to convolutional layers would enable end-to-end QUBO-based training, though the parameter coupling structure of convolutional filters differs from fully-connected layers. Finally, incorporating line search, acceptance thresholds, or trust-region constraints could improve the fraction of beneficial updates (currently 54--57\% of iterations reduce the loss, the complement of the 43--46\% that temporarily increase it as noted above).


\section*{Acknowledgments}
M. Atallah acknowledges support from the 2025 SandboxAQ Research Excellence Scholarship. M. Atallah and R. Herrman acknowledge University of Tennessee Knoxville AI Tennessee seed funding. The funder played no role in study design, data collection, analysis and interpretation of data, or the writing of this manuscript.

\section*{Competing interests}
All authors declare no financial or non-financial competing interests.

\section*{Code and Data Availability}
\label{sec:codeAndDataAvail}
The code and data for this research can be found at \url{https://github.com/Mostafa-Atallah2020/qml-qubo}.

\input{layerwise_gram_qubo.bbl}

\end{document}